\documentclass{ifacconf}
\usepackage{textcomp}
\usepackage{amsmath,amssymb,amsfonts}%
\usepackage{gensymb}
\usepackage{booktabs}
\usepackage{chemformula}
\usepackage{eurosym}
\usepackage{url}
\usepackage{graphicx}      
\usepackage{natbib}        

\begin{document}
\begin{frontmatter}

\title{GreenLight-Gym: Reinforcement learning benchmark environment for control of greenhouse production systems}
\thanks[footnoteinfo]{This is an accepted pre-print version for the 8th IFAC Conference on Sensing, Control and Automation Technologies for Agriculture, AgriControl 2025.}

\author[First]{Bart van Laatum} 
\author[First]{Eldert J. van Henten} 
\author[Second]{Sjoerd Boersma}

\address[First]{Agricultural Biosystems Engineering, Wageningen University\\ (e-mail: bart.vanlaatum@wur.nl)}
\address[Second]{Biometris, Wageningen University \& Research}

\begin{abstract}                
This study presents GreenLight-Gym\footnote{https://github.com/BartvLaatum/GreenLight-Gym}, a new, fast, open-source benchmark environment for developing reinforcement learning (RL) methods in greenhouse crop production control. Built on the state-of-the-art GreenLight model, it features a differentiable C++ implementation leveraging the CasADi framework for efficient numerical integration. GreenLight-Gym improves simulation speed by a factor of 17 over the original GreenLight implementation. A modular Python environment wrapper enables flexible configuration of control tasks and RL-based controllers. This flexibility is demonstrated by learning controllers under parametric uncertainty using two well-known RL algorithms. GreenLight-Gym provides a standardized benchmark for advancing RL methodologies and evaluating greenhouse control solutions under diverse conditions. The greenhouse control community is encouraged to use and extend this benchmark to accelerate innovation in greenhouse crop production.
\end{abstract}


\end{frontmatter}

\section{Introduction}
Greenhouse production systems allow for year-round food production by providing resilience against outdoor weather events, which is much needed given the increased number of extreme weather events due to climate change \citep{fuExtremeRainfallReduces2023}. In contrast to open-field agriculture, greenhouse production systems offer more facets to control the crop’s environment to maintain an ideal climate for crop growth and quality \citep{stanghelliniGreenhouseHorticultureTechnology2019}. In practice, experienced growers manually manage the greenhouse climate by relying on their expertise and visual observations of the crops. However, the sector struggles to find skilled growers capable of efficiently and manually controlling the greenhouse climate \citep{sparksWhatCurrentState2018}. Moreover, expanding facilities impedes manual climate management.

Given these challenges, advanced control methods are required to support or even automate the grower's decision-making to control more large-scale greenhouse systems with more efficiency and minimal labor. Advanced control methods such as model predictive control (MPC) optimize the greenhouse operations by repeatedly solving a finite horizon optimal control problem \citep{vanhentenGreenhouseClimateManagement1994, vanstratenOptimalControlGreenhouse2010}. However, MPC methods demand high computational efforts, which becomes more pressing for high-fidelity models and long-term horizons. Especially under system uncertainty due to sensor inaccuracies and modeling errors \citep{boersmaRobustSamplebasedModel2022,kuijpersWeatherForecastError2022}.



More recently, reinforcement learning (RL) has emerged as a promising framework for developing control policies, i.e., controllers, for greenhouse production systems \citep{zhangRobustModelbasedReinforcement2021, morcegoReinforcementLearningModel2023}. Contrasting MPC, RL learns (near-)optimal control policies offline from numerous interactions with a simulation model. This enables it to learn how uncertainty affects control performance using stochastic simulations during training without requiring real-time optimization.

However, no standardized simulation environments exist to benchmark RL-based control methods in greenhouse production control. Without a standardized benchmark, it is difficult to assess the progress of RL-based methodologies for greenhouse control and determine the direction in which they should evolve. Early studies applied RL to simplified process-based greenhouse models \citep{morcegoReinforcementLearningModel2023}, but these models do not match the dynamics of today's high-tech greenhouses. Other studies used the high-fidelity greenhouse simulator KASPRO to identify a greenhouse model and learn a controller with model-based reinforcement learning \citep{zhangRobustModelbasedReinforcement2021,caoIGrowSmartAgriculture2022}. Although KASPRO accurately models real-world greenhouse dynamics \citep{dezwartAnalyzingEnergysavingOptions1996}, its closed-source nature makes it unsuitable as a standardized RL benchmark environment.

Several high-fidelity simulation models have enhanced the accuracy of open-source greenhouse models \citep{altes-buchGreenhousesModelicaLibrary2019, katzinGreenLightOpenSource2020,qiuGreenLightPlus2024}. However, this higher fidelity comes at the expense of simulation speed, which hinders training data-intensive RL controllers. Additionally, these open-source simulation models lack the flexibility to diversify the training data, which is required to prevent RL from overfitting \citep{kirkSurveyZeroshotGeneralisation2023}. Training data diversity can, for instance, be achieved by adding parametric uncertainty, leading to randomized model dynamics. Or by presenting weather trajectories from a wide range of climates. Tantamount the above, existing greenhouse simulators face challenges regarding computational speed, open-source, and flexibility, even though these are essential elements for an RL benchmark training environment.

This work addresses these challenges by introducing GreenLight-Gym, a new open-source simulation environment to benchmark RL-based control methods for greenhouse crop production systems. GreenLight-Gym reimplements the state-of-the-art greenhouse model GreenLight \citep{katzinGreenLightOpenSource2020} in Python and C++ using the CasADi framework \citep{anderssonCasADiSoftwareFramework2019}. CasADi significantly speeds up simulations for GreenLight facilitating the rapid development of RL-based controllers. GreenLight-Gym comes with a modular Python environment wrapper that can provide users with the flexibility to define the training environment. For instance, by randomizing GreenLight's model parameters or weather trajectories. This characteristic benefits the development of RL-based controllers that demonstrate the generalizing capabilities to unseen greenhouse circumstances.

This paper presents GreenLight-Gym as an open-source RL benchmark for greenhouse production, comparing its computational efficiency with existing GreenLight implementations. We then demonstrate its use by training two well-known RL controllers under parametric uncertainty. Section \ref{sec:greenlight-gym} describes GreenLight-Gym’s implementation and modular components, Section \ref{sec:formalisation-rl} formalizes the RL framework using GreenLight-Gym, Section \ref{sec:rl-det-exp} outlines the simulation experiments, and Section \ref{sec:results} presents results. Following this initial release, further GreenLight-Gym extensions are planned. This paper concludes by outlining the development roadmap.


\section{GreenLight-Gym}\label{sec:greenlight-gym}
\begin{figure}
    \centering
    \includegraphics[width=\linewidth]{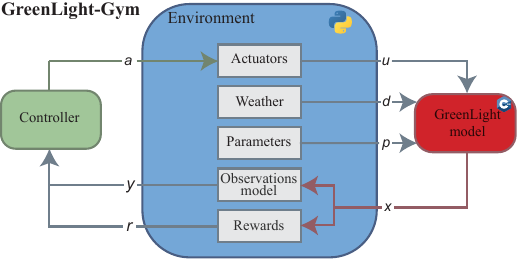}
    \caption{Architecture of GreenLight-Gym. The controller generates an action $a$, which is applied to the GreenLight-Gym simulation environment. The next state $x$ is computed using a CasADi-based implementation that simulates greenhouse dynamics, incorporating controllable inputs and external disturbances. The resulting state is used to compute the reward function and update the controller.}
    \label{fig:gl-gym-architecture}
\end{figure}

GreenLight-Gym was designed with the following distinctive features in mind: (1) \textbf{High simulation speed} to rapidly develop RL-based controllers on a high-fidelity greenhouse model, and (2) \textbf{Modularity} to empower users with full control over greenhouse simulation settings facilitating reproducible and comparable simulation studies. Figure \ref{fig:gl-gym-architecture} overviews GreenLight-Gym's architecture, illustrating how the controller interacts with the simulation model. The remainder of this section details the implementation of the GreenLight model, the environment wrapper, and the controllers.

\subsection{GreenLight model}\label{sec:greenlight-model} 
GreenLight-Gym is built upon GreenLight, a state-of-the-art, process-based greenhouse model for production simulations in high-tech greenhouse facilities \citep{katzinGreenLightOpenSource2020}.
The model incorporates supplementary lighting to capture its influence on the greenhouse climate and crop by extending \citet{vanthoorMethodologyModelbasedGreenhouse2011a} climate model and employing a simplified version of \citet{vanthoorMethodologyModelbasedGreenhouse2011b} tomato yield model. Since GreenLight does not model a realistic energy management system--which typically includes a combined heat and power unit, a boiler, and a heat buffer--this work assumes that resources such as energy and \ch{CO2} are directly purchased from the market.

The non-linear model has been discretized and written in standardized state-space form:
\begin{equation}\label{eq:greenlightmodel}
    x_{k+1} = f(x_k, u_k, d_k, p),
\end{equation}
where $k\in \mathbb{Z}^{\geq0}$ represents the discrete time step, $x\in \mathbb{R}^{28}$ is the state of the greenhouse model consisting of indoor climate and crop variables, $u\in \mathbb{R}^{8}$ is the controllable input that affects the indoor climate, $d\in \mathbb{R}^{10}$ is the non-controllable input, i.e., weather disturbance and $p\in\mathbb{R}^{208}$ represents the model parameters. For a detailed model description, we refer to \citep{katzinGreenLightOpenSource2020}.

The GreenLight model has been implemented in C++ using the symbolic framework from CasADi \citep{anderssonCasADiSoftwareFramework2019}. CasADi supports implicit numerical integration methods and automatic differentiation to approximate the continuous model dynamics. Therefore, after compilation, CasADi can calculate the value of the Jacobian using a simple function evaluation rather than intensive approximation methods. Additionally, the GreenLight-Gym environment uses CasADi's just-in-time (JIT) compilation capabilities. JIT compilation transforms the discrete-time model \eqref{eq:greenlightmodel} into efficient C code, giving significant gains in computational performance. GreenLight-Gym offers the interface to call the JIT compiled model function from Python with (non-)controllable inputs and model parameters, after which the state at the next time step is returned.

\subsection{Environment}\label{sec:modularity}
The GreenLight-Gym environment is organized into distinctive modules to preserve flexible and modular software. This gives users complete control over the simulation characteristics, allowing them to diversify the training environment and reproduce and compare RL-based control methodologies. 

\subsubsection{Weather} 
The weather module is designed to integrate weather data from various years and geographic locations. This feature enables the generation of a fixed benchmark environment to assert control performance under comparable weather conditions. Moreover, it facilitates research to study the generalization capabilities of RL controllers across geographic locations.

\subsubsection{Parameters} GreenLight-Gym includes user-configurable model parameters that enable the specification of a wide range of greenhouse constructions and tomato crop varieties. This allows the greenhouse control community to benchmark performance across diverse greenhouse and tomato species. Moreover, diversifying model parameters during training across varied crop parameters mitigates uncertainty and can enhance transfer to real-world greenhouses \citep{kirkSurveyZeroshotGeneralisation2023}. 




\subsubsection{Observation model} GreenLight-Gym provides a modular configuration of the observation model. Several observation types include indoor climate and crop state, control input, time, outdoor climate, and weather forecasts. This modular design of the observations allows users to define a greenhouse control problem where RL controller inputs match real-world greenhouse sensors and measurements. Additionally, the module enables users to encode extra information in the observation space as a function of model state $x$, control input $u$, disturbance input $d$, and model parameters $p$.

\subsubsection{Rewards} 
 This module allows users to encode desired objectives in a real-valued reward function, balancing control costs with crop-growth benefits while penalizing unfavorable states. A mismatch between the model and the real-world system in greenhouse production control could yield high rewards in unfavorable states when purely focusing on profitability. For instance, when the adverse effects of extreme humidity concentrations on crop yield and quality are neglected. GreenLight-Gym's reward module facilitates reward function design to encode desired controller behavior by combining one or multiple reward components that assign an immediate value to an action in a specific state.

\subsection{Controller}\label{sec:controller}
The controller module within GreenLight-Gym was designed to support RL-based and rule-based controllers. This design allows users to implement and evaluate different types of controllers, providing a standardized framework for testing and comparison. 

The RL-based controllers are implemented using the Stable Baselines3 library \citep{raffin}. The GreenLight-Gym contains two well-known RL algorithms: Proximal Policy Optimization (PPO) and Soft-Actor Critic (SAC), but users can also customize RL algorithms with Stable Baselines3. All the algorithm's hyperparameters can be specified in a YAML configuration file. Additionally, hyperparameter sweeps are supported using the Weights and Biases platform\footnote{\url{https://wandb.ai/}}.

Additionally, implementing rule-based controllers allows users to encode general grower control strategies. These rule-based controllers can serve as a baseline for RL-based controllers. The following key points, derived from \citep{katzinEnergySavingsGreenhouses2021}, depict the implemented rule-based controller: Lamps $(u_{\text{lamp}})$ are on between 00:00-18:00, unless outdoor radiation exceeds $400$ W m$^{-2}$, or the predicted outdoor global solar radiation sum over the day is above $10\text{ MJ m}^{-2}\text{ d}^{-1}$. CO$_2$ $(u_{\text{CO}_2})$ is injected during the light period whenever the concentration drops below 800 ppm (P-controlled). Heating $(u_{\text{boil}})$ is applied whenever the indoor temperature drops below the target set-point, $19.5\degree$C during the light period or $16.5\degree$C during the dark period (P-controlled). Roof ventilation windows $(u_{\text{vent}})$ are opened whenever the indoor temperature is $5\degree$C above the target set point or if the indoor relative humidity is over 85\%. Vents are closed if the indoor temperature is 1 \degree C below the target set point (both P-controlled). Thermal screens $(u_{\mathrm{thScr}})$ are closed if the outdoor temperature is below $5\degree$C during the day and below $10\degree$C at night. Screens are opened if the indoor temperature is $4\degree$C above the target set-point or when the indoor relative humidity is over 85\%. Blackout screens $(u_{\mathrm{blScr}})$ are closed if the lamps are on during the dark period. These control rules can be altered in a configurable YAML file if desired.

The subsequent section formalizes the greenhouse control problem for RL using the components introduced in the latter section.

\section{Reinforcement learning for greenhouse production control}\label{sec:formalisation-rl}
\subsection{RL formulation}
We employ an RL-based controller to optimize greenhouse production by manipulating climatic actuators. The controller architecture is represented by a parameterized deep neural network $(\pi_{\boldsymbol{\theta}})$, which maps observations $y_k$ to actions $a_k$. The GreenLight-Gym actuator module maps this action to control input $u_k$ and queries the GreenLight model to compute the next state of the system $x_k$. Finally, the state is used to calculate the following observation and reward $r_k$, which are both used by RL to optimize its parameterized controller $\pi_{\theta}$.

The following equation estimates the next state and observation of the system:
\begin{align}\label{eq:observation-model}
    \hat{x}_{k+1} &= f(\hat{x}_k, u_k, d_k, \hat{p}_k)\nonumber \\
    \hat{y}_k &= g(\hat{x}_k, u_k, d_k, \hat{p}_k),
\end{align}
where the non-linear $f(\cdot)$ function and its input arguments are similar defined as in \eqref{eq:greenlightmodel}. $g(\cdot)$ represents the user defined observation model, see subsection \ref{sec:modularity}. The $\hat{\cdot}$ operator indicates that the variable is an estimation of the true variable. 

This study uses $\hat{p}_k$ to represent the stochastic parameter variable and defines it as:
\begin{align}\label{eq:phat_epsilon_dist}
    \hat{p}_k &= \mu_p(1+\epsilon),\qquad \epsilon \sim \mathcal{U}(-\delta, \delta),
\end{align}
where $\mu_p$ represents the nominal parameter values, and $\epsilon$ follows the given uniform distribution, where $\delta\in[0,1)$ represents the uncertainty range. Such that $\hat{p}\in[\mu_p(1-\delta), \mu_p(1+\delta)]$, and $\mathbb{E}[\hat{p}]=\mu_p$ ensures positive parameter values.

Additionally, the controller receives a custom reward $r_k$. This reward is a numerical feedback signal that indicates how well the action in the previously observed state was in terms of the desired outcome. The goal of RL is to find a (near-)optimal controller that maximizes the expected cumulative reward over time:
\begin{equation}\label{eq:RL-objective}
     \pi_{\boldsymbol{\theta}}^{*}=\arg \max_{\boldsymbol{\theta}}J(\boldsymbol{\theta})=\mathbb{E}_{\tau\sim\rho(\tau|\pi_{\boldsymbol{\theta}}, \hat{p})} \left[\sum_{k=0}^{\infty} \gamma^k r_k(y_k, u_k)\right],
\end{equation}
where $\pi_{\boldsymbol{\theta}}^{*}$ denotes the optimal controller, and $\gamma\in[0,1]$ discounts future rewards. Trajectory $\tau$ represents a feasible sequence of states and controls: $(x_0, u_0, \dots,x_N, u_N)$.
$\tau\sim\rho(\tau|\pi_{\boldsymbol{\theta}}, \hat{p})$ represents the distribution of trajectories $\tau$ given controller $\pi_{\boldsymbol{\theta}}$ and sampled parameters $\hat{p}$.

\subsection{Reward function}
The objective for greenhouse crop production control is to maximize the greenhouse's operational return, i.e., economic performance indicator (EPI), while satisfying system constraints. Since RL does not natively deal with state constraints, these are encoded in the reward function. Therefore, the function consists of two components: encoding EPI and penalizing constraint violations.

The EPI was defined by revenue, driven by tomato fruit growth, minus resource consumption, resulting from boiler activation, CO$_2$ injection, and turning on the lights. The EPI reward function denotes as:
\begin{align*}\label{eq:profit-reward}
    r_k^{\text{EPI}}(y_k, u_k)& = \\
    c_{\text{frt}} \Delta y_{k,\text{frt}} - & r 
    (c_{\mathrm{CO}_2} u_{k,\mathrm{CO}_2}+c_{\mathrm{boil}} u_{k,\mathrm{boil}}+ c_{\mathrm{lamp}} u_{k,\mathrm{lamp}}),
\end{align*}
where $\Delta y_{k, \mathrm{frt}}$ indicates the incremental fruit growth over the last time step, and $u_{k, i}$ represents the resource consumption of actuator $i$. The cost coefficients $c_{i}$ are defined for \ch{CO2} injection, boiler activation, and lighting. The electric cost for managing $u_{\mathrm{thScr}}$,  $u_{\mathrm{vent}}$,  $u_{\mathrm{blScr}}$ are negligible, and thus deliberately omitted in the EPI reward function. 

State constraints were encoded via linear a penalty function. The linear penalty function is defined as follows:
\begin{equation}\label{eq:pen-reward}
    r^{\text{pen}}_{k}(y_k) = \sum_i^{N_{cs}} P_{k, i}(y_{k, i}),
\end{equation}
where $P_{k, i}$ denotes the penalty value for state variable $i$ at time step $k$, and $N_{cs}$ denotes the number of state constraints. The penalty value is formally denoted as:
\begin{equation}
    P_{k, i}(y_{k,i}) =     
    \begin{cases}
      y_{k,i} - y_{\mathrm{max}, i} & \text{if } y_{k, i} > y_{\mathrm{max}, i}, \\
      y_{\mathrm{min}, i} - y_{k,i} & \text{if } y_{k, i} < y_{\mathrm{min}, i}, \\
      0 & \text{otherwise.}
    \end{cases}       
\end{equation}

The final reward combines the EPI reward and the linear penalty function as follows:
\begin{equation}\label{eq:additive-rew}
    r_k(y_k, u_k)= r^{\text{EPI}}_k(y_k, u_k)-r^{\text{pen}}_k(y_k).
\end{equation} 

The EPI reward is scaled between $[0, 1]$ to stabilize the learning process. The minimum value was defined as having no harvest and maximum resource consumption. Having the maximum possible harvest without using resources defined the maximum value. Additionally, to prevent the penalty value from dominating the cumulative reward estimates in \eqref{eq:RL-objective}, each penalty value $P_{k, i}$ was min-max scaled between $[0,1]$. The min and maximum value for each $P_{k, i}$ was manually determined. 

\section{Simulation methodology}\label{sec:experiments}
This work conducted two simulation experiments illustrating GreenLight-Gym's capabilities for accelerating simulation, benchmarking RL controllers, and handling parametric uncertainties. First, the simulation speed of GreenLight-Gym was compared against two implementations of GreenLight. Second, two well-known RL algorithms, PPO and SAC, trained a controller under parametric uncertainty. The controllers were benchmarked against the rule-based baseline implemented in GreenLight-Gym.


\subsection{Simulation Speed}
Since accelerated simulation speed is a critical feature for RL training, the execution time of GreenLight-Gym was compared against the original implementation of GreenLight (GL-Matlab) and a recent implementation of GreenLight in Python (GL-Python) \citep{katzinGreenLightOpenSource2020, qiuGreenLightPlus2024}. The simulation speeds were evaluated using a desktop with a 12-core Intel(R) Xeon(R) W-2133 CPU at up to 3.60GHz. All three implementations used identical weather disturbances $d$ and control inputs $u$ from a prior validation experiment \citep{katzinGreenLightOpenSource2020}. Each simulation spanned 10 growing days, starting in October 2009. The time discrete model from \eqref{eq:greenlightmodel} had a step size of $\Delta t = 300(\mathrm{s})$. For each simulation run, the steps per second were calculated.

\subsection{RL for greenhouse production control under parametric uncertainty}\label{sec:rl-det-exp}
GreenLight-Gym provides a user-friendly approach for training controllers using two off-the-shelf RL algorithms, PPO and SAC, on challenging greenhouse control problems. This simulation experiment tests the effect of uncertainty stemming from variations in crop model parameters on the performance of RL-based controllers. The RL-based controllers were trained using Stable Baselines3. A rule-based baseline controller was simulated for comparison purposes. The goal of the controllers was maximizing EPI while minimizing state constraint violations over 60 consecutive days. 

The RL-based controllers attempted to solve the optimization problem formulated in \eqref{eq:RL-objective} with stochastic crop parameter values. During training, 28 crop parameters were randomized using \eqref{eq:phat_epsilon_dist} at each time step by sampling $\epsilon$ from the distribution defined in \eqref{eq:phat_epsilon_dist}. The other model parameter values equal $\mu_p$. Seven uncertainty values $(\delta)$ were analyzed, ranging from 0\% to 30\% of the nominal crop parameter values. The reward function $r_k$ was defined as in \eqref{eq:additive-rew}. Using the following price parameters were used: $c_{\mathrm{frt}} = 1.6 \text{~\euro{} kg/m}^2$, $c_{\mathrm{CO_2}}=0.3 \text{~\euro{} kg/m}^2$, $c_{\mathrm{boil}}=0.09 \text{~\euro{} W/m}^2$, $c_{\mathrm{lamp}}=0.3 \text{~\euro{} W/m}^2$. The constraints on the three indoor climate variables (Temperature; \ch{CO2}-concentration; Relative humidity) were defined as:
\begin{equation}
    y_{\mathrm{min}} = \text{(15 300 50)}^{\mathrm{T}}, y_{\mathrm{max}} = \text{(34 1600 85)}^{\mathrm{T}}
\end{equation}
The controllers were trained to control six inputs that manipulate the indoor climate: $u_{\mathrm{boil}}, u_{\mathrm{CO_2}},u_{\mathrm{thScr}}, u_{\mathrm{vent}}, u_{\mathrm{lamp}},$ and $u_{\mathrm{blScr}}$, using the following bounds:
\begin{equation}    
u_{\mathrm{min}} =  \text{(0 0 0 0 0 0)}^{\mathrm{T}}, u_{\mathrm{max}} =  \text{(130 5.0 1 1 116 1)}^{\mathrm{T}}
\end{equation}

Table \ref{tab:observation-space} lists the controller inputs. 




         

The used weather trajectory $d$ spans from 1 March to 30 April 2010, recorded at Schiphol Airport. The weather remained fixed during training and testing. Although this weather scenario does not reflect real-world variability, it demonstrates GreenLight-Gym's capabilities in developing various RL-based controllers. Future work should address generalization to new weather trajectories. PPO and SAC were trained using separate neural networks for the actor and the critic, each consisting of a multilayer perceptron (MLP) with three hidden layers. The actor network used 256 nodes, while the critic had 512. The algorithms' hyperparameters were informed through a random search. Table \ref{tab:ppo-sac-hyperparameters} summarizes the hyperparameter values. Unreported hyperparameters used Stable Baselines3's default values.

\begin{table}[ht]
    \centering
    \caption{Observation inputs to the controllers.}
    \begin{tabular}{l|l|l}
         \toprule
         \textbf{Symbol} & \textbf{Description} & \textbf{Unit} \\
         \hline
         $y_{\mathrm{T}}$          & Air temperature           & $\degree$C \\
         $y_{\ch{CO2}}$            & \ch{CO2} concentration    & ppm \\
         $y_{\mathrm{RH}}$         & Relative humidity         & \% \\
         $y_{\mathrm{tPipe}}$      & Pipe temperature          & $\degree$C \\
         $y_{\mathrm{CFrt}}$       & Fruit dry weight          & kg/m$^2$ \\
         $y_{\mathrm{t24Crop}}$    & Daily avg. canopy temp.   & $\degree$C \\
         $y_{\mathrm{tCanSum}}$    & Canopy temp. sum          & $\degree$C \\
         
         \hline
         $y_{\mathrm{Hour_{sin,cos}}}$ & Encoded hour of the day & - \\
         $y_{\mathrm{Day_{sin,cos}}}$  & Encoded day of the year & - \\
         
         \hline
         $u_{\mathrm{boil}}$       & Boiler activation         & W/m$^2$ \\
         $u_{\mathrm{CO_2}}$       & \ch{CO2} injection        & mg/m$^2$/s\\
         $u_{\mathrm{thScr}}$      & Thermal screen closure    & - \\
         $u_{\mathrm{vent}}$       & Roof ventilation opening  & - \\
         $u_{\mathrm{lamp}}$       & Lamp electrical input     & W/m$^2$ \\
         $u_{\mathrm{blScr}}$      & Blackout screen closure   & - \\
         
         \hline
         $d_{\mathrm{iGlob}}$      & Global radiation          & W/m$^2$ \\ 
         $d_{\mathrm{T}}$          & Outdoor temperature       & $\degree$C \\ 
         $d_{\mathrm{RH}}$         & Outdoor relative humidity & \% \\ 
         $d_{\ch{CO2}}$            & Outdoor \ch{CO2} concentration & ppm \\ 
         $d_{\mathrm{Wind}}$       & Wind speed                & m/s \\
         
         \bottomrule
    \end{tabular}
    \label{tab:observation-space}
\end{table}
\begin{table}[ht]
    \centering
    \caption{Hyperparameters for PPO and SAC. Abbreviations: fn = function, Ent coef = entropy coefficient, VF = value function.}
    \begin{tabular}{l|c|c}
        \toprule
        \textbf{Hyperparameter} & \textbf{PPO} & \textbf{SAC} \\
        \midrule
        Total timesteps              & 2M         & 2M \\
        Learning rate $(\alpha)$                & $2 \times 10^{-5}$ & $7 \times 10^{-4}$ \\
        Batch size                   & 128        & 128 \\
        Discount factor $(\gamma)$                     & 0.9631     & 0.9631 \\
        Activation fn               & SiLU          & SiLU\\
        \midrule
        $n_{\text{steps}}$           & 2048       & - \\
        $n_{\text{epochs}}$          & 8          & - \\
        GAE $\lambda$                & 0.9167     & - \\
        Clip range                   & 0.2        & - \\
        Ent coef                     & 0.05434    & - \\
        VF coef                      & 0.8225     & - \\
        \midrule
        Buffer size                  & -          & 576.1K \\
        Learn starts                 & -          & 57.6K \\
        Polyak-coef $(\tau)$         & -          & 0.0135 \\
        Train freq                   & -          & 50 \\
        Grad steps                   & -          & 10 \\
        \bottomrule
    \end{tabular}
    \label{tab:ppo-sac-hyperparameters}
\end{table}

The cumulative reward $\sum_{k=0}^{N}r_{k}(y_k, u_k)$, cumulative EPI $\sum_{k=0}^{N}r_{k}^{\mathrm{EPI}}(y_k, u_k)$, and cumulative penalty $\sum_{k=0}^{N}r_{k}^{\mathrm{pen}}(y_k, u_k)$ evaluated controller performance. Each controller was evaluated by taking the average over 30 simulation runs to account for variability in the outcomes due to parametric uncertainty.

\section{Simulation Results \& Discussion}\label{sec:results}

\subsection{Simulation speed}
GreenLight-Gym demonstrates significant improvement in the simulation speed over other implementations of GreenLight. GreenLight-Gym can simulate over 1800 steps per second on a single CPU core, offering speedups of $\sim$24 and $\sim$17 over the Matlab-based and Python-based GreenLight implementation, respectively. This makes collecting over a million data points possible within ten minutes. The implementation of GreenLight-Gym has been validated against the original Matlab code and is available upon request.


\subsection{RL for greenhouse production control}
\begin{figure*}[t]
    \centering
    \includegraphics[width=\textwidth]{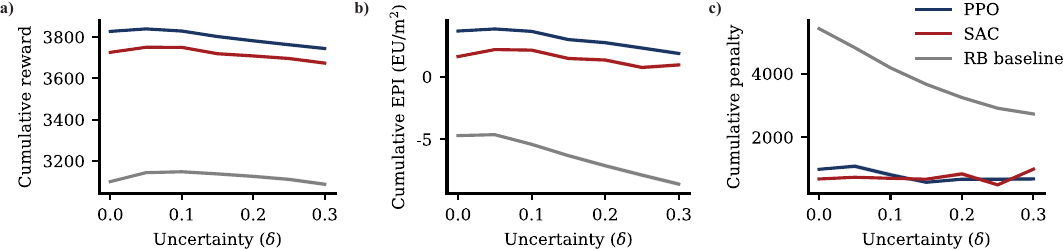}
    \caption{Controller performance under parametric uncertainty. The mean performance metric for PPO, SAC, and rule-based (RB) controllers at varying parametric uncertainty levels averaged over 30 simulation runs.}
    \label{fig:uncertainty}
\end{figure*}

Figure \ref{fig:uncertainty}.a illustrates controller performance under parametric crop model uncertainty. RL-based controllers outperform the rule-based baseline across the entire range of uncertainty values. Introducing uncertainty up to 10\% improves the performance of all three controllers, possibly due to the beneficial parameter variations that lead to increased crop growth. The increased EPI reflects this finding at these low uncertainty levels. This effect drops after further increasing uncertainty, and the performance trends decrease.

Despite the decrease in EPI, observed in Figure \ref{fig:uncertainty}.b, at higher noise levels, both PPO and SAC remain substantially above zero across the entire uncertainty range. By contrast, the RB baseline starts around $-5\text{\EUR{}{}/m}^2$ and becomes progressively worse. These results demonstrate that the learned controllers can better maintain positive economic performance under increasing parameter uncertainty, whereas the fixed rule‐based strategy suffers from performance degradation. This underperformance in terms of EPI could be due to the high cost of electricity. The RL controllers effectively learned that the high electricity cost of supplementary lighting outweighed its revenue benefits and thus minimized its usage.

Randomizing crop model parameters had low effects on the penalties for state constraint violations for both RL-based controllers. The cumulative penalty for PPO and SAC barely fluctuates across the entire uncertainty range. The limited effect of crop parameters on the climate dynamics could declare these results. The rule-based baseline reports significantly higher penalties for state constraint violations. This could result from the delayed effect of opening the roof ventilation on humidity. Conversely, the RL controllers can handle these delayed responses. Tuning the control rules from subsection \ref{sec:controller} would yield a more competitive baseline.

The reported controller performance should be interpreted as an upper limit since accurate input measurement and complete knowledge of the weather trajectory were assumed. Moreover, the model's simplified dynamics may not fully capture the complexities of real-world greenhouse systems. Nevertheless, studying advanced control methods in other complex greenhouse control tasks using GreenLight-Gym could support transferring these findings to these systems. Overall, both RL-based controllers outperform the rule‐based approach, highlighting their robustness in maintaining performance across varying degrees of uncertainty.

\section{Conclusion \& future work}
This work presented GreenLight-Gym as a benchmark environment for developing RL-based controllers for greenhouse production systems. The environment increased simulation speed over the original GreenLight implementation by a factor of 17. Additionally, the environment modular implementation facilitates customizing greenhouse simulation and RL controller settings. Its flexibility as a benchmark environment was demonstrated by evaluating RL controller performance under parametric uncertainty and comparing them against a rule-based baseline. GreenLight-Gym's open-source implementation aims to stimulate the development of innovative RL-based control methods for greenhouse systems. 



Future work can go in two directions: extending the environment or developing more advanced control methods for greenhouse control. Potential GreenLight-Gym extensions include, but are not limited to, adding different crop model species to develop general RL controllers that transfer to varying crops; integrating more realistic energy management systems to better reflect real-world greenhouse operations; include photo-realistic image generation models for greenhouse crops to add morphological information to RL-based controllers. 

Future studies could include MPC formulations for more state-of-the-art baselines for RL or investigate the integration of RL and MPC. Moreover, further simulation studies could assert RL's robustness against uncertainties arising from unseen weather trajectories, sensor inaccuracies, or modeling errors.


\bibliography{ifacconf}             
\end{document}